\documentclass[useAMS,usenatbib,onecolumn]{mn2e}
\usepackage{amsmath}
\usepackage{epsfig,color}
\usepackage{dcolumn}
\usepackage[utf8]{inputenc}
\usepackage{graphics,graphicx}
\usepackage{epstopdf}
\usepackage{float}
\usepackage[para]{threeparttable}
\usepackage{url}
\usepackage{color}

\newcommand{\cm}{cm$^{-1}$}
\newcommand{\icm}{cm\textsuperscript{-1}}
\newcommand{\hhoo}{H$_{2}$O$_2$}

\newcommand{\ai}{\textit{ab initio}}

\newcommand{\Dh}[1]{${\mathcal D}_{#1{\rm h}}$}

\title[ExoMol line lists XV: H$_2$O$_2$]{ExoMol line lists XV: A new hot line list for hydrogen peroxide}

\date{\today}

\author[Al-Refaie et al]{Ahmed F. Al-Refaie$^{1}$, Oleg
L. Polyansky$^{1,2}$,  Roman I. Ovsyannikov$^{2}$,
\newauthor Jonathan Tennyson$^{1}$ and Sergei N. Yurchenko$^{1}$,\\
$^{1}$Department of Physics and Astronomy, University College London, Gower Street,
WC1E 6BT London, UK\\
$^{2}$Institute of Applied Physics, Russian Academy of
Sciences, Ulyanov Street 46, Nizhny Novgorod, Russia 603950}

\date{Accepted XXXX. Received XXXX; in original form XXXX}

\pagerange{\pageref{firstpage}--\pageref{lastpage}} \pubyear{2016}

\begin{document}

\label{firstpage}

\maketitle

\begin{abstract}
A computed line list for hydrogen peroxide, H$_2{}^{16}$O$_2$, applicable to
temperatures up to $T=1250$~K is presented.  A semi-empirical high
accuracy potential energy surface is  constructed
and used with an {\it  ab initio} dipole moment surface as input TROVE to compute
7.5 million rotational-vibrational states and
around 20 billion transitions with
associated Einstein-$A$ coefficients for rotational excitations up to
$J=85$. The resulting APTY line list is complete
for wavenumbers below 6~000 cm$^{-1}$ ($\lambda < 1.67$~$\mu$m) and
temperatures up to 1250~K.
Room-temperature spectra are compared with
laboratory
measurements and data currently available in the HITRAN database and literature.
Our rms with line positions from the literature is 0.152 \cm\
and our absolute intensities agree better than 10\%.  The full line list is
available from the CDS database as well as at \url{www.exomol.com}.
\end{abstract}


\section{Introduction}


Terrestrial hydrogen peroxide exists as a trace molecule in the Earth's
atmosphere and contributes to the atmospheres oxidising budget as well as ozone
production and water chemistry \citep{74DaXXXX.H2O2,91ChJoTr.H2O2,13AlAbBe.H2O2,14ZiKrxx.H2O2}
and its concentration is now being routinely observed \citep{13AlAbBe.H2O2}.
Astrophysically there have been multiple
detections of \hhoo\ in the atmosphere of
Mars \citep{04ClSaMo.H2O2,04EnBeGr.H2O2,12EnGrLe.H2O2,15AoGiKa.H2O2} with
seasonal variation, possibly formed by triboelectricity in dust devils and dust
storms \citep{12EnGrLe.H2O2} and may well act as an agent in the oxidization of
the Martian surface. Hydrogen peroxide has also been detected in the atmosphere of
Europa \citep{13HaBrXX.H2O2} in the 3.5$~\mu m$ region. The first detection of
interstellar \hhoo\ was made by \citet{11BePaLi.H2O2} and is believed to
play an important role in astrophysical water chemistry similar to that on
Earth. \citet{12DuPaBe.H2O2} suggest that \hhoo\ is produced on
dust-grains via the hydrogenation of grain HO$_{2}$ and released into the gas-phase
through surface reactions. On the dust-grain, \hhoo\ acts as an intermediate in the
formation of water and aids in the production of other species such
as  H$_2$CO, CH$_3$OH, and O$_2$.


Hydrogen peroxide belongs to the peroxide group of molecules with an HO-OH bond
dissociation enthalpy of 17050 \cm \citep{96BaAySc.H2O2} at 0 K. \hhoo\ is an asymmetric prolate rotor
molecule and is the simplest molecule that exhibits internal rotation. This
torsional motion gives rise to a double minimum potential curve with respect to
its internal rotation co-ordinates as well as two alignments of the O-H bonds:
\textit{cis} and \textit{trans}. The consequence of this motion means that there
are four sub-levels for each torsional excitation which are characterized by
their symmetry. This necessitates the use of an additional quantum number, $\tau$, to
unambiguously describe its motion. The molecular states can be classified using the $C^{+}_{\rm
2h}$(M) symmetry group which best describes the torsional splitting caused by
the \textit{cis} and \textit{trans} tunneling \citep{84HoXXXX.H2O2}. \hhoo\ has
six vibrational modes:  $\nu_{1}$ and $\nu_{5}$ represent the symmetric and
asymmetric O-H stretching respectively, $\nu_{3}$ and $\nu_{6}$ represent the
O-H bending modes, $\nu_{2}$ represents the O-O stretch and the $\nu_{4}$
mode represents the torsional excitation with the more common notation of $n$.

Experimental studies of ro-vibrational \hhoo\ spectra
have mostly probed the torsional motion in the ground \citep{88OlHuYo.H2O2}, the $\nu_{3}$ \citep{92CaFlJo.H2O2} and
$\nu_{6}$ \citep{90PeFlCa.H2O2,95PeVaFl.H2O2} vibrational modes. Conversely, the higher-lying O-H stretching modes,
$\nu_{1}$ and $\nu_{5}$, are poorly studied using high resolution techniques. The difference between the two
stretching bands is about 8 -- 10 \icm\ and torsional splitting from the double minimum
of
the potential gives rise to doubling \citep{74GiSrXX.H2O2} in the form of
'quasi'-degenerate states \citep{09RaKnWe.H2O2} that are difficult to resolve
with a degree of accuracy. \citet{88OlHuYo.H2O2} give an estimate of 3610
- 3618~\icm\ for $\nu_{5}$ and 3601 - 3617~\icm\ for $\nu_{1}$ whilst a Raman
study gives a lower value of 3607~\icm\  for the
~$\nu_{1}$~ band-centre \citep{74GiSrXX.H2O2} but determining the accuracy to better
than 0.1~\icm\ is difficult.

\hhoo\ has been a
benchmark system for developing
methods aiming to treat large amplitude motion
\citep{00LuXXXX.H2O2,02MlXXXX.H2O2,02YuMuXX.H2O2,09CaHaBo.H2O2}.
Recent
calculations on the ro-vibrational states for \hhoo\ include the {\it ab initio}
computation using  CCSD(T)-F12 electronic structure calculations
of band frequencies accurate to about 4.0 \icm\ by
\citet{09RaKnWe.H2O2},
models of the peroxide stretches by \citet{05BaBiXX.H2O2}, a discrete variable representation (DVR) calculation
for levels up to 6000 \icm\ by \citet{01ChMaGu.H2O2,03LiGuxx.H2O2} and finally, a potential energy surface
(PES) calculations  by \citet{98KoCaHa.H2O2} and \citet{99KuRiLu.H2O2}. Calculation which also consider
transition intensities are rather rarer but a recent example is provided
by  \citet{11CaShBo.H2O2}. 
The peroxide system was used to benchmark the large amplitude calculations of
MULTIMODE \citep{03BoCaHa.methods} up to $J=20$ and showed good agreement
against HITRAN line intensities but the PES used had an rms of $\approx$ 20
\icm\ against experimental band centers.
However, this PES has been superceded by the higher accuracy \ai\
potential energy surface (PES) of \citet{13MaKoXX.H2O2}
which was further modified by \citet{jt553}. This modified PES was
used for our room-temperature line list \citep{jt620} and provides
the starting point for the refinements performed here.

Experimental transitions frequencies and  intensities  for \hhoo\ are
available in the HITRAN 2012 database \citep{jt557} but only for room temperature modelling up to 1800~\cm.
This region covers the torsional, O-H bending modes
and O-O stretch but misses the O-H stretches in the 3750~\cm~region. Only a
few studies deal with absolute intensities of \hhoo\ in the far-infrared
\citep{41ZuGiXX.H2O2,96PeFlCA.H2O2,95PeVaFl.H2O2} with only PNNL-IR \citep{PNNL}
data providing integrated intensities in the mid-infrared region
\citep{09JoSaBu.H2O2}. The thermal decomposition of hydrogen peroxide at
423 K makes it difficult and dangerous to study at higher temperatures.

Theoretical line lists  can be used to fill in
gaps in the experimental data both in terms of wavelength and temperature coverage. The ExoMol
project \citep{jt528} aims to produce comprehensive theoretical molecular line
lists to aid in studies of the atmospheres of exoplanets,
cool stars and other (hot) bodies. A room temperature line list for \hhoo\ was
previously computed by us \citep{jt620} using the PES of
\citet{jt553} and a new \ai\ dipole moment surface (DMS). It provides
about 1 billion transitions at up to 8,000~\icm. However it is limited as the
rotational excitation of $J=40$ makes it inadequate for high temperature
modelling and the lower energy cut-off means that coverage above 4,000~\icm\
rapidly becomes incomplete. This work aims to build upon this line list by
refining the PES towards spectroscopic accuracy and extending the temperature
and frequency range for which the resulting line list is applicable.

\section{Method}
\label{s:method}

\subsection{Potential energy Surface Refinement}
Our previous room-temperature  \hhoo\ line list \citep{jt620} was computed
using the \ai\ PES of \citet{13MaKoXX.H2O2} with the small
adjustment of the
\ai\ equilibrium geometry and height of the torsional  barrier proposed
by \citet{jt553}. This PES reproduces the known empirical
energy levels with a root mean square (rms) of about 1--2 \icm. However,
by utilising empirical band-centre shifting \citep{jt466} during the
computation of the Hamiltonian, this was significantly reduced  it to 0.001 -- 0.1~\icm\ in
the room temperature line list. The empirical shifting can be
thought of as an addition to the \ai\ PES producing an 'ad-hoc' PES which we
will refer to as 'H2O2-2015' in comparisons below. Whilst band-centre shifting can reproduce
experimental energies simply, its accuracy and predictive ability is limited to
vibrational bands whose band centre positions are already experimentally characterised.
A more robust  method of correcting the PES is through fitting or refining to experimental
energies. The TROVE nuclear motion program used by us here, see below,
provides PES refinement capabilities and has been successful in
producing accurate PES for molecules such as H$_2$CO \citep{11YaYuJe.H2CO},
NH$_3$ \citep{jt500}, SO$_3$ \cite{jt580}, PH$_3$ \citep{jt592}  and CH$_4$\citep{jt564}.

The procedure implemented in TROVE describes a correction surface $\Delta V$ to the initial
(\ai) surface $V$. The new refined surface $V'$ can therefore be written as $V' =
V + \Delta V$ and the new Hamiltonian as $H' = H + \Delta V$, where $H$ is the
Hamiltonian if the starting point for the refinement.
The wavefunctions from $H$, are used as basis-functions for $H'$.
\(\Delta V\) is expanded in Taylor series and the expansion coefficients, \(\Delta
f_{ijklmn}\),  are obtained in a variational least-squares fit to spectroscopic
data via the objective function $F$:
\begin{equation}
 F = \sum_{i}^{N}w_{i}(E_{i}^{obs} - E_{i})^{2} = 0 ,
\end{equation}
where $N$ is the number of observed energies $E_{i}^{obs}$, $E_{i}$ are
the calculated energies and $w_{i}$ are the weights.

Here the original 282 expansion coefficients of the \ai\ PES are reduced to 163
by removing the symmetry-related O-H stretching ($ij$) and bending ($kl$) terms from the
input PES and simply linking them in the computation of potential energy
terms in the Hamiltonian. This ensures that the symmetry of these terms is
preserved  during the fitting process.
The quality of the fit is determined by the quality and vibrational diversity of
the input dataset. Two sources of experimental data come from line-positions
provided in the literature and transitions from HITRAN.
The HITRAN dataset sources come from observations by
\citet{95PeVaFl.H2O2}, \citet{96PeFlCA.H2O2}, \citet{90PeFlCa.H2O2} and \citet{Klee1999154} with literature line-positions from \citet{89FlCaJo.H2O2}, \citet{88OlHuYo.H2O2} \citet{50GiXXXX.H2O2}, \citet{41ZuGiXX.H2O2} and
\citet{92CaFlJo.H2O2}. This empirical dataset provides the $\nu_{4}$,
$\nu_{3}+\nu_{4}$, $\nu_{4}+\nu_{6}$ and $\nu_{2}$ vibrational terms.
Unfortunately there is little to no reliable data on the $\nu_{1}$ and $\nu_{5}$ energy levels, their
reported band-centre values vary significantly in literature making them unsuitable
for the refinement. This hampers the vibrational diversity that would aid in construction of an extensive fitted PES.
However, these terms can be indirectly improved by including
higher $J$ values from other vibrational states.

Our input dataset includes all energies for $J\leq4$ up to 4000~\icm. The
weights $w_{i}$ used have an arbitrary range of values that are normalized in the fit.
The energies given in literature are the simplest to include in the refinement process and are given the highest weighting. Here the pure torsional band at $J>0$ from \citet{92CaFlJo.H2O2} and \citet{88OlHuYo.H2O2} are given the highest weighting of $w_{i}=100$. The $\nu_2$, $\nu_3$, $\nu_6$, $\nu_3+\nu_4$, $\nu_4 + \nu_6$ energies and \hhoo\ band centers (except for $\nu_1$ and $\nu_5$) from \citet{92CaFlJo.H2O2}, \citet{50GiXXXX.H2O2}, \citet{90PeFlCa.H2O2}, \citet{89FlCaJo.H2O2} and \citet{41ZuGiXX.H2O2} are given weights $10\leq w_{i}\leq20$.

Transitions from HITRAN require additional work. In order to determine the upper state of a transition requires the assignment of the
lower state energy. Fortunately HITRAN provides the lower state energy for all
transitions in the database. However, lower state energies require corroboration from literature data and/or the \ai\ energies for the upper state energies to be included in the fit with $1\leq w_{i}\leq 9$ based on confidence of the datum.
Each input datum must be correlated with a theoretically computed
energy level which, in this present work, was straightforward due to the
good agreement given by the initial \ai\ PES.

Special measures must be taken in order to ensure that the refinement process does not lead to unphysical shapes for the new PES due to a limited sampling of the experimental data not covering all the complexity of the potential energy surface of HOOH. For example the high stretching or bending overtones are poorly represented in the experimental and therefore it is important to retain the \ai\ quality of the original PES by \citet{jt553}. To this end we constrain the PES around \ai\ energies at each geometry \citep{03YuCaJe.PH3,jt503,11YaYuJe.H2CO,jt592}.

The new potential energy surface is called H2O2-2016. Table
\ref{tab:Weights-rms} describes the rms for states of a particular weight. The high 
quality of the H2O2-2016 energies computed without any empirical band shifts 
shows that this new semi-empirical PES performs better overall than the \ai\
band-shifted PES especially for the lower weighted states. Weights $\geq$10
relate to vibrational states that were involved in the band-shifting which gives H2O2-2015
its low rms values. Comparing weights lower than 10 suggests that the
predictive ability of the H2O2-2016 PES is greatly enhanced. The overall comparison as a function of the rotational quantum number $J$ with a weighted rms is given in Table
\ref{tab:J-N-fit-rms}.

\begin{table}
\centering
\caption{Comparison of $N$ weighted experimental data-points in the fit and
non-weighted root mean squared deviation of both H2O2-2016 (this work) and
H2O2-2015 \citep{jt620} for each dataset. }
\begin{tabular}{lrlll}
\hline\hline
  &     & H2O2-2016   & H202-2015   \\
Weight & N   & rms (\icm) & rms (\icm) & Comment \\
\hline
100 & 43  & 0.001 & 0.000 & $J>0$ pure torsional states\\
10-20 & 144  & 0.004 & 0.007 & Band centers and $\nu_3+\nu_4$, $\nu_6+\nu_4$
states\\
1-9 & 186  & 0.539 & 1.369 & Upper state extracted from HITRAN with corroborated lower states\\
\hline
\end{tabular}
\label{tab:Weights-rms}
\end{table}
\begin{table}
\centering
\caption{Comparision of $N$ experimental data-points in the fit and weighted
root mean squared deviation of both H2O2-2016 (this work) and H2O2-2015
\citep{jt620}. }
\begin{tabular}{llll}
\hline\hline
  &     & H2O2-2016   & H202-2015   \\
J & N   & wrms (\icm) & wrms (\icm)  \\
\hline
0 & 34  & 0.238 & 0.254 \\
1 & 47  & 0.079 & 0.320 \\
2 & 81  & 0.096 & 0.345 \\
3 & 116 & 0.183 & 0.404 \\
4 & 132 & 0.154 & 0.287 \\
Total & & 0.150 & 0.321 \\
\hline
\end{tabular}
\label{tab:J-N-fit-rms}
\end{table}

\begin{table}
\centering
\caption{Residuals in \icm\ for energies computed from the H2O2-2016 PES. Observed data is from \citet{92CaFlJo.H2O2} and \citet{90PeFlCa.H2O2}. The overall rms is 0.0642 \icm}
\label{tab:residuals}
\begin{tabular}{llllllrrr|llllllrrr}
\hline\hline
$J$  & $K$ & $\nu_3$ & $n$ & $\tau$ & $\nu_6$ & Obs     & Calc     & O-C & $J$  & $K$ & $\nu_3$ & $n$ & $\tau$ & $\nu_6$ & Obs  & Calc  & O-C \\
\hline
30 & 0  & 0 & 0 & 1 & 0 & 789.58  & 789.64  & -0.06 & 31 & 11 & 0 & 0 & 2 & 0 & 1950.60 & 1950.55 & 0.05  \\
30 & 1  & 0 & 0 & 2 & 0 & 793.05  & 793.12  & -0.07 & 31 & 11 & 0 & 0 & 3 & 0 & 1963.12 & 1963.07 & 0.05  \\
30 & 2  & 0 & 0 & 1 & 0 & 829.29  & 829.34  & -0.05 & 31 & 5  & 0 & 1 & 2 & 1 & 2577.85 & 2577.91 & -0.06 \\
30 & 2  & 0 & 0 & 4 & 0 & 841.34  & 841.32  & 0.02  & 31 & 1  & 0 & 2 & 2 & 1 & 2694.38 & 2694.48 & -0.10 \\
30 & 3  & 0 & 0 & 2 & 0 & 876.03  & 876.07  & -0.04 & 31 & 2  & 0 & 2 & 1 & 1 & 2729.02 & 2729.07 & -0.05 \\
30 & 3  & 0 & 0 & 2 & 0 & 876.03  & 876.07  & -0.04 & 31 & 3  & 0 & 2 & 2 & 1 & 2775.74 & 2775.66 & 0.08  \\
30 & 4  & 0 & 0 & 1 & 0 & 940.03  & 940.07  & -0.04 & 31 & 9  & 0 & 0 & 3 & 1 & 2877.92 & 2877.84 & 0.08  \\
30 & 6  & 0 & 0 & 4 & 0 & 1134.61 & 1134.58 & 0.03  & 31 & 7  & 0 & 1 & 3 & 1 & 2940.33 & 2940.25 & 0.08  \\
30 & 2  & 0 & 1 & 4 & 0 & 1198.50 & 1198.58 & -0.08 & 31 & 9  & 0 & 1 & 3 & 1 & 3232.19 & 3232.21 & -0.02 \\
30 & 7  & 0 & 0 & 2 & 0 & 1241.30 & 1241.33 & -0.03 & 32 & 1  & 0 & 0 & 2 & 0 & 898.70  & 898.78  & -0.08 \\
30 & 7  & 0 & 0 & 2 & 0 & 1241.30 & 1241.33 & -0.03 & 32 & 2  & 0 & 0 & 1 & 0 & 936.15  & 936.21  & -0.06 \\
30 & 7  & 0 & 0 & 3 & 0 & 1253.29 & 1253.25 & 0.04  & 32 & 2  & 0 & 0 & 4 & 0 & 948.32  & 948.29  & 0.02  \\
30 & 4  & 0 & 1 & 4 & 0 & 1308.34 & 1308.43 & -0.09 & 32 & 3  & 0 & 0 & 2 & 0 & 983.10  & 983.16  & -0.05 \\
30 & 1  & 0 & 2 & 2 & 0 & 1362.75 & 1362.83 & -0.08 & 32 & 4  & 0 & 0 & 1 & 0 & 1047.03 & 1047.07 & -0.04 \\
30 & 8  & 0 & 0 & 4 & 0 & 1390.06 & 1390.01 & 0.05  & 32 & 2  & 0 & 1 & 4 & 0 & 1305.21 & 1305.30 & -0.09 \\
30 & 5  & 0 & 1 & 3 & 0 & 1390.16 & 1390.27 & -0.11 & 32 & 7  & 0 & 0 & 2 & 0 & 1347.93 & 1347.96 & -0.03 \\
30 & 2  & 0 & 2 & 1 & 0 & 1396.42 & 1396.52 & -0.10 & 32 & 1  & 0 & 2 & 2 & 0 & 1468.52 & 1468.61 & -0.09 \\
30 & 9  & 0 & 0 & 3 & 0 & 1544.80 & 1544.75 & 0.05  & 32 & 8  & 0 & 0 & 4 & 0 & 1496.82 & 1496.77 & 0.05  \\
30 & 0  & 1 & 0 & 1 & 0 & 1645.15 & 1645.20 & -0.05 & 32 & 9  & 0 & 0 & 3 & 0 & 1651.43 & 1651.37 & 0.06  \\
30 & 2  & 1 & 0 & 2 & 0 & 1688.11 & 1688.15 & -0.03 & 32 & 0  & 1 & 0 & 1 & 0 & 1749.83 & 1749.89 & -0.06 \\
30 & 10 & 0 & 0 & 1 & 0 & 1707.36 & 1707.33 & 0.03  & 32 & 1  & 1 & 0 & 2 & 0 & 1752.86 & 1752.92 & -0.07 \\
30 & 10 & 0 & 0 & 4 & 0 & 1717.16 & 1717.10 & 0.06  & 32 & 2  & 1 & 0 & 1 & 0 & 1794.43 & 1794.46 & -0.03 \\
30 & 3  & 1 & 0 & 1 & 0 & 1731.55 & 1731.59 & -0.04 & 32 & 10 & 0 & 0 & 4 & 0 & 1823.55 & 1823.48 & 0.07  \\
30 & 11 & 0 & 0 & 2 & 0 & 1898.17 & 1898.11 & 0.06  & 32 & 3  & 1 & 0 & 2 & 0 & 1837.22 & 1837.26 & -0.04 \\
30 & 11 & 0 & 0 & 3 & 0 & 1910.47 & 1910.41 & 0.06  & 32 & 5  & 0 & 1 & 2 & 1 & 2631.92 & 2632.00 & -0.08 \\
30 & 6  & 1 & 0 & 2 & 0 & 1978.18 & 1978.19 & 0.00  & 32 & 9  & 0 & 0 & 3 & 1 & 2931.74 & 2931.70 & 0.04  \\
30 & 5  & 0 & 1 & 2 & 1 & 2525.44 & 2525.50 & -0.06 & 32 & 9  & 0 & 1 & 3 & 1 & 3285.71 & 3285.72 & -0.01 \\
30 & 0  & 0 & 2 & 1 & 1 & 2638.21 & 2638.21 & 0.00  & 33 & 4  & 0 & 0 & 1 & 0 & 1103.04 & 1103.09 & -0.05 \\
30 & 2  & 0 & 2 & 1 & 1 & 2678.98 & 2678.96 & 0.02  & 33 & 7  & 0 & 0 & 2 & 0 & 1403.73 & 1403.76 & -0.03 \\
30 & 6  & 0 & 1 & 4 & 1 & 2768.27 & 2768.35 & -0.08 & 33 & 9  & 0 & 0 & 3 & 0 & 1707.24 & 1707.18 & 0.06  \\
30 & 9  & 0 & 0 & 3 & 1 & 2825.75 & 2825.74 & 0.01  & 33 & 10 & 0 & 0 & 4 & 0 & 1879.23 & 1879.16 & 0.07  \\
30 & 9  & 0 & 1 & 3 & 1 & 3180.32 & 3180.35 & -0.03 & 33 & 5  & 0 & 1 & 2 & 1 & 2687.68 & 2687.77 & -0.09 \\
30 & 11 & 0 & 0 & 3 & 1 & 3196.72 & 3196.74 & -0.02 & 33 & 7  & 0 & 2 & 2 & 1 & 3252.13 & 3252.10 & 0.03  \\
30 & 10 & 0 & 1 & 4 & 1 & 3355.50 & 3355.61 & -0.11 & 33 & 9  & 0 & 1 & 3 & 1 & 3340.86 & 3340.86 & 0.00  \\
31 & 3  & 0 & 0 & 2 & 0 & 928.92  & 928.97  & -0.04 & 34 & 2  & 0 & 0 & 4 & 0 & 1061.96 & 1061.93 & 0.03  \\
31 & 4  & 0 & 0 & 1 & 0 & 992.69  & 992.73  & -0.04 & 34 & 3  & 0 & 0 & 2 & 0 & 1096.87 & 1096.93 & -0.06 \\
31 & 6  & 0 & 0 & 4 & 0 & 1187.24 & 1187.21 & 0.03  & 34 & 4  & 0 & 0 & 1 & 0 & 1160.72 & 1160.77 & -0.05 \\
31 & 1  & 0 & 1 & 3 & 0 & 1230.77 & 1230.86 & -0.09 & 34 & 2  & 0 & 1 & 4 & 0 & 1418.58 & 1418.69 & -0.11 \\
31 & 2  & 0 & 1 & 4 & 0 & 1253.76 & 1253.86 & -0.10 & 34 & 7  & 0 & 0 & 2 & 0 & 1461.18 & 1461.23 & -0.05 \\
31 & 7  & 0 & 0 & 2 & 0 & 1293.78 & 1293.81 & -0.03 & 34 & 10 & 0 & 0 & 4 & 0 & 1936.55 & 1936.48 & 0.07  \\
31 & 7  & 0 & 0 & 2 & 0 & 1293.78 & 1293.81 & -0.03 & 34 & 7  & 0 & 2 & 2 & 0 & 2026.09 & 2026.08 & 0.01  \\
31 & 0  & 0 & 2 & 1 & 0 & 1410.64 & 1410.73 & -0.09 & 34 & 0  & 0 & 0 & 4 & 1 & 2287.94 & 2288.05 & -0.11 \\
31 & 0  & 0 & 2 & 1 & 0 & 1410.64 & 1410.73 & -0.09 & 34 & 5  & 0 & 1 & 2 & 1 & 2745.11 & 2745.21 & -0.10 \\
31 & 8  & 0 & 0 & 4 & 0 & 1442.61 & 1442.56 & 0.05  & 34 & 9  & 0 & 1 & 3 & 1 & 3397.65 & 3397.63 & 0.02  \\
31 & 9  & 0 & 0 & 3 & 0 & 1597.29 & 1597.23 & 0.06  & 35 & 1  & 0 & 0 & 2 & 0 & 1091.72 & 1091.77 & -0.05 \\
31 & 1  & 1 & 0 & 2 & 0 & 1717.38 & 1717.60 & -0.23 & 35 & 4  & 0 & 0 & 1 & 0 & 1220.07 & 1220.13 & -0.06 \\
31 & 2  & 1 & 0 & 1 & 0 & 1736.75 & 1736.79 & -0.05 & 35 & 7  & 0 & 0 & 2 & 0 & 1520.29 & 1520.34 & -0.05 \\
31 & 10 & 0 & 0 & 4 & 0 & 1769.53 & 1769.46 & 0.07  & 35 & 1  & 0 & 0 & 2 & 1 & 2324.23 & 2324.16 & 0.07  \\
31 & 3  & 1 & 0 & 2 & 0 & 1783.76 & 1783.79 & -0.03 & 35 & 1  & 0 & 0 & 2 & 1 & 2324.23 & 2324.16 & 0.07  \\
31 & 4  & 1 & 0 & 1 & 0 & 1847.46 & 1847.49 & -0.03 & 36 & 2  & 0 & 0 & 4 & 0 & 1182.24 & 1182.21 & 0.03 \\
\hline
\end{tabular}
\end{table}

Overall H2O2-2016 improves the rms deviations of H2O2-2015 by more than a factor of 2. Table \ref{tab:residuals} highlights residuals for $J\geq30$
for the $\nu_3$ and $\nu_6$ line positions from \citet{92CaFlJo.H2O2} and \citet{90PeFlCa.H2O2} and shows excellent agreement with an overall rms of 0.064 \cm.
The rms deviation for all 2734 states in HITRAN up to $J=49$ and energy up to
3461.02~\icm\ is 0.834~\icm. Vibrational terms that correspond to the highest weighted states have an rms
of 0.192~\icm. Around 12 states related to higher excited torsional modes $n>3$
have an rms of 5.2~\icm\ and may well be due to misassignments.
This PES is therefore of there of improved accuracy and is the one used below.
The coefficients defining this PES are given in the Supplementary Information.

\subsection{Variational computation}

Theoretical ROVibrational Energies (TROVE) \citep{07YuThJe.method} was employed
to compute the ro-vibrational energies of \hhoo. TROVE is a variational
nuclear motion solver and was used to successfully produce the room temperature
\hhoo\ line list as well as hot line lists for NH$_3$ \citep{jt500}, CH$_{4}$
\citep{jt564}, PH$_{3}$ \citep{jt592}, H$_{2}$CO \citep{jt597} and SO$_3$ \citep{jt641}.

TROVE can operate with any co-ordinate system of our choosing by utilizing an
approximate kinetic energy operator (KEO). For \hhoo, this approximate KEO was shown
by \citet{jt553} to
produce results that are in very good agreement with the exact KEO code
WAVR4 \citep{jt339}. However, the computational cost is greatly reduced for rotationally excited states
by using TROVE.
Convergence of the KEO usually requires an
expansion order of 6 or 8 \citep{07YuThJe.method}; 6 being chosen for this
work and 8 for the potential energy expansion as suggested from a previous \hhoo\ calculation by \citet{jt553}.

Basis sets and wave functions are symmetrized to the \Dh{2}(M) molecular symmetry group
\citep{98BuJexx.method} which is isomorphic to the $C^{+}_{\rm 2h}$(M) symmetry group.
This has the benefit of factorizing the Hamiltonian into smaller independent blocks for diagonalization.
The irreducible representations of this group are A$_{g}$, A$_{u}$, B$_{1g}$,
B$_{1u}$, B$_{2g}$, B$_{2u}$, B$_{3g}$ and B$_{3u}$. However, the states
corresponding to B$_{2g}$, B$_{2u}$, B$_{3g}$ and B$_{3u}$ have zero nuclear statistical
weight and therefore these matrix blocks are not needed for $J>0$. The corresponding $J=0$
energies from these representations are also unphysical, but computed and kept
as vibrational band centres for reference purposes.

A symmetry-adapted basis-set is constructed by a multi-step contraction scheme
by solving the 1D Schr\"{o}dinger equation via the Numerov-Cooley
method \citep{24Nuxxxx.method,61Coxxxx.method} for the basis-functions
$\phi_{n_{i}}(\zeta_{i}) (i=1,2,...,6)$, where $\zeta_{1}$ 
represents the O-O stretching co-ordinate, $\zeta_{2})$ and $\zeta_{3}$ represent the O-H
stretching co-ordinates, $\zeta_{4}$ and $\zeta_{5}$ represents the
O-H bending modes and  $\zeta_{6}$ represents the torsional mode.
$n_{i}$ is the local mode quantum number assigned by TROVE. The basis functions $\phi_{n_{i}}$ 
are then used to form a product-type basis set, which is truncated by the polyad number $P$:
\begin{equation}
\label{eq:polay}
  P = 4n_{1} + 8(n_{2}+n_{3}+n_{4}+n_{5})+n_{6} \leq P_{\rm max}.
\end{equation}
We use $P_{\rm max}=42$ as it was found to give good convergence \citep{jt553}.
The second step requires computing a contracted basis set by reducing the six
dimensional co-ordinates into four dimensions. This accomplished by reducing the
6-dimensional co-ordinate system into four subspaces: $(\zeta_{1})$,
$(\zeta_{2},\zeta_{3})$, $(\zeta_{4},\zeta_{5})$ and
$(\zeta_{6})$ based on their permutation properties. The reduced Hamiltonian is solved using the
primitives $\phi_{n_{i}}$ as basis functions to obtain the symmetrised
vibrational eigenfunctions $\Phi_{\lambda_1}(\zeta_{1})$,
$\Phi_{\lambda_2}(\zeta_{2},\zeta_{3})$, $\Phi_{\lambda_3}(\zeta_{4},\zeta_{5})$ and
$\Phi_{\lambda_4}(\zeta_{6})$. The vibrational contracted basis set is then formed
from symmetrized products of these eigenfunctions which is also truncated via Eq.~(\ref{eq:polay}).
A final contraction step is performed by solving the $J=0$ problem
and replacing the bulky primitive and contracted vibrational basis set
with the more compact $J=0$ wavefunctions $\Psi^{J=0,\Gamma}_{i}$ truncated at
12~000~\icm. The benefit of this is that
the computation of the Hamiltonian matrix elements for $J>0$ is more efficient
and reduces the size of the matrix. Whilst this form means we can replace the diagonal vibrational terms with experimental band centers, the low rms error of H2O2-2016 PES band centers makes this procedure unnecessary.
Our final ro-vibrational wavefunction has the form:
\begin{equation}
\Psi^{J,\Gamma} = \sum_{i,K,p}c_{i,K,p}\Psi^{J=0,\Gamma}_{i}| J,K,p\rangle,
\label{eq:rovib}
\end{equation}
where $| J,K,p\rangle$ are rigid rotor functions with parity $p$ defined
by \citet{05YuCaJe.NH3} and $c_{i,K,p}$ are the eigenvector coefficients
obtained by diagonalization of the Hamiltonian matrix. The linear
algebra libraries LAPACK \citep{99AnBaBi.method} and SCALAPACK \citep{slug}
were employed to solve for the eigenvalues and eigenvectors.

TROVE assigns six local mode $n_{i}$ quantum numbers to each state by finding
the largest contributing $c_{i,K,p}$. These can be reassigned to match the more
standard
normal modes $v_{i}$ generally used in literature. However, TROVE only
provides six quantum numbers for
reassignment, when seven are required by the inclusion of $\tau$.
The $\tau$ quantum number can be preserved in the reassignment in
TROVE by utilizing the following form:
\begin{equation}
  v_{4} = 4n + i,
\end{equation}
where $n$ is the excitation and $i$ is the symmetry where $i=0,1,2,3$ is
A$_{g}$, B$_{2u}$, B$_{2g}$ and A$_{u}$ respectively. To retrieve $n$ and $\tau$
simply requires:
\begin{equation}
  \tau = (v_{4} \bmod 4) + 1 , \quad n =
\left\lfloor\frac{v_{4}}{4}\right\rfloor .
\end{equation}


\subsection{Dipole moment surface  and intensities }

An electric dipole moment surface (DMS) is required in order to compute
transition intensities. The \ai\ DMS of \citet{jt620} was utilized; this was computed at the CCSD(T)-f12b \citep{07BaMuxx.ai} level of theory in the frozen-core
approximation and is applicable to energies up to $hc$~$12~000$~\icm.

The eigenvectors, obtained by diagonalization, are  used in
conjunction with the DMS to compute the required linestrengths and
Einstein-$A$ coefficients of
transitions that satisfy the rotational selection rules
\begin{equation}\label{e:rot-rules}
   J^{\prime} - J^{\prime\prime} = 0,\pm 1, J^{\prime} + J^{\prime\prime} \ne 0
\end{equation}
and the symmetry selection rules:
\begin{equation}\label{e:sym-rules}
 A_{g} \leftrightarrow A_{u}, B_{1g} \leftrightarrow B_{1u}
\end{equation}
applied to para and ortho states, respectively.
States with $B_{2g}, B_{2u}, B_{3g}$ and $B_{3u}$ symmetry are forbidden due their zero
nuclear statistical weights.

The Einstein-$A$ coefficient for a particular transition from the
\textit{initial} state $i$ to the \textit{final} state $f$ is given by:
\begin{equation}
A_{if} = \frac{8\pi^{4}\tilde{\nu}_{if}^{3}}{3h}(2J_{i} + 1)\sum_{A=X,Y,Z}
|\langle \Psi^{f} | \bar{\mu}_{A} | \Psi^{i} \rangle |^{2},
\end{equation}
where $J_{i}$ is the rotational quantum number for the initial state, $h$ is
Planck's constant, \(\tilde{\nu}_{if}\) is the transition frequency (\(hc \,
\tilde{\nu}_{if} = E_{f} -E_{i}\)), \(\Psi^{f}\) and \(\Psi^{i}\) represent the
eigenfunctions of the final and initial states respectively, \(\bar{\mu}_{A}\)
is the
electronically averaged component of the dipole moment along the space-fixed
axis \(A=X,Y,Z\) (see also \citet{05YuThCa.method}). From this the absolute
absorption intensity is determined by:
\begin{equation}
\nonumber
I(f \leftarrow i) = \frac{A_{if}}{8\pi c}g_{\rm ns}(2
J_f+1)\frac{\exp\left(-\frac{c_2 \tilde{E}_{i}}{T}\right)}{Q\;
\tilde{\nu}_{if}^{2}}\left[1-\exp\left(\frac{-c_2\tilde{\nu}_{if}}{T}
\right)\right],
\end{equation}
where \(c_2 ={hc}/{k}\) is the second radiation constant, $\tilde{E}_i$ is the energy term value,
\(T\) the absolute temperature and \(g_{\rm ns}\)
is the nuclear spin statistical weight factor. \(Q\), the partition function, is
given
by:
\begin{equation}
  Q = \sum_{i} g_{i}\exp\left({\frac{-c_2 \tilde{E}_{i}}{T}}\right),
  \label{eq:part}
\end{equation}
where \(g_{i}\) is the degeneracy of a particular state \(i\) with the energy
term value \(\tilde{E}_{i}\). For
\hhoo, \(g_{i}\) is \(g_{\rm ns}(2 J_i + 1)\) with $g_{\rm ns} = 1$ for
\(A_{g}\) and
\(A_{u}\) symmetries and $g_{\rm ns} = 3$  for \(B_{1g}\) and \(B_{1u}\)
symmetries.  The
transitions were computed using the energy limits $hc$ 6\,000 and $hc$
12\,000~\cm\ for
the lower and upper states, respectively giving a complete coverage of the
region 0 -- 6\,000~\cm\ ($\lambda > 1.67$~$\mu$m).

The intensities were computed using an enhanced version of TROVE that utilizes
the nVidia graphics processing units (GPU) allowing for the computation of
5,000--30,000 transitions per second on a single GPU. The GPUs utilized were the
nVidia M2090, K20 and the K40 models.
A paper discussing this will be published elsewhere \citep{jtGAIN}.

\section{Results}

\begin{figure}
\centering
\epsfxsize=12.0cm \epsfbox{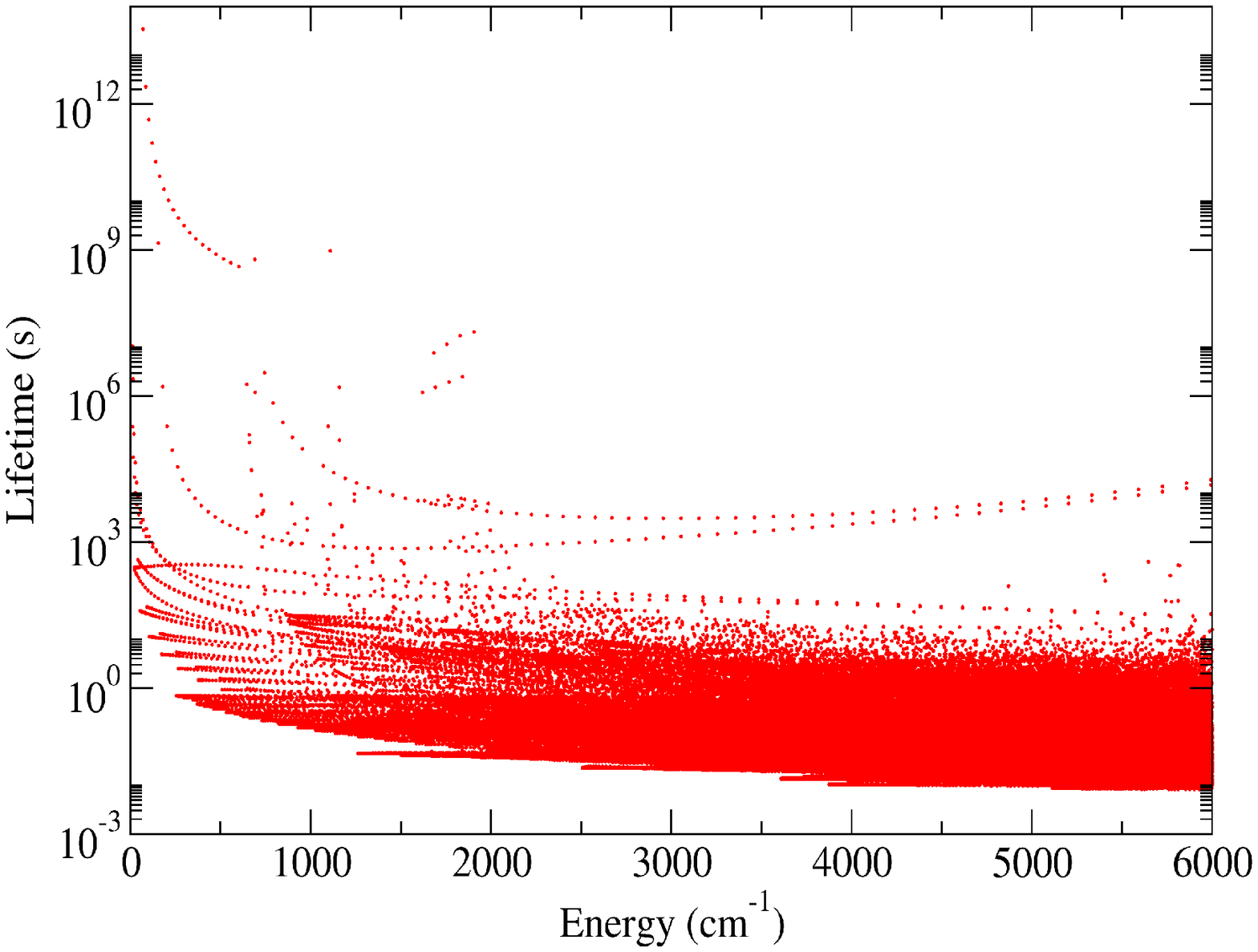}
\caption{Radiative lifetimes computed \citep{jt624} for \hhoo\ states up to 6,000 \icm}
\label{fig:H2O2_lifetimes}
\end{figure}

The final hot line list named APTY contains 7~560~352 states  and  almost 20
billion transitions that completely cover the 0 -- 6\,000~\icm\ region.
An extended line list is provided which contains an additional 8 billion
transitions in the 6\,000-- 8\,000~\icm\ region with reduced completeness
for higher temperature. Radiative lifetimes are also computed via the methodology presented by \citet{jt624}. Figure  \ref{fig:H2O2_lifetimes} presents the lifetimes computed for states up to 6\,000 \icm. Here the complicated rotational structure of \hhoo\ gives rise to long lived lower state energies with lifetimes that can reach up to 30 million years.
Extract from the ExoMol-form \citep{jt548,jt631}  states and transition files are given
in Tables \ref{tab:levels} and \ref{tab:trans}.
Spectra at arbitrary
temperatures can be computed using the Einstein-$A$ coefficients from the
transition files.

\begin{table}
\caption{Extract from the \hhoo\ state file. The full table is available at
\protect\url{http://cdsarc.u-strasbg.fr/cgi-bin/VizieR?-source=J/MNRAS/}.
} \label{tab:levels}
\begin{center}
\footnotesize
\tabcolsep=5pt
\begin{tabular}{lrccccccccccccccccccccccc}
\hline
$I$  &  \multicolumn{1}{c}{$\tilde{E}$ ,\cm}   &  $g$  &  $J$ & $L$ &
$\Gamma_{\rm tot}$ & $v_1$ & $v_2$ & $v_3$ & $v_4$ & $v_5$ & $v_6$ &
$\tau$ & $\Gamma_{\rm vib}$ & $K$ & $p$  & $\Gamma_{\rm rot}$ & $ I_{J,\Gamma}$
& $|C_i^{2}|$ & $n_1$ & $n_2$ & $n_3$ & $n_4$ & $n_5$ & $n_6$ \\
\hline
1  & 0.0         & 1 & 0 & inf & 1 & 0 & 0 & 0 & 0 & 0 & 0 & 1 & 1  & 0 & 0 & 1 & 1  & 0.95 & 0 & 0 & 0 & 0 & 0 & 0  \\
2  & 254.54376   & 1 & 0 & 6.9464D-01 & 1 & 0 & 0 & 0 & 1 & 0 & 0 & 1 & 1  & 0 & 0 & 1 & 2  & 0.94 & 0 & 0 & 0 & 0 & 0 & 4  \\
3  & 569.726553  & 1 & 0 & 3.1256D-01 & 1 & 0 & 0 & 0 & 2 & 0 & 0 & 1 & 1  & 0 & 0 & 1 & 3  & 0.94 & 0 & 0 & 0 & 0 & 0 & 8  \\
4  & 865.942367  & 1 & 0 & 3.1317D+01 & 1 & 0 & 0 & 1 & 0 & 0 & 0 & 1 & 1  & 0 & 0 & 1 & 4  & 0.87 & 1 & 0 & 0 & 0 & 0 & 0  \\
5  & 1000.872029 & 1 & 0 & 1.7264D-01 & 1 & 0 & 0 & 0 & 3 & 0 & 0 & 1 & 1  & 0 & 0 & 1 & 5  & 0.93 & 0 & 0 & 0 & 0 & 0 & 12 \\
6  & 1119.712431 & 1 & 0 & 6.8795D-01 & 1 & 0 & 0 & 1 & 1 & 0 & 0 & 1 & 1  & 0 & 0 & 1 & 6  & 0.85 & 1 & 0 & 0 & 0 & 0 & 4  \\
7  & 1391.868926 & 1 & 0 & 6.8101D+00 & 1 & 0 & 1 & 0 & 0 & 0 & 0 & 1 & 1  & 0 & 0 & 1 & 7  & 0.44 & 0 & 0 & 0 & 1 & 0 & 0  \\
8  & 1432.296785 & 1 & 0 & 2.9134D-01 & 1 & 0 & 0 & 1 & 2 & 0 & 0 & 1 & 1  & 0 & 0 & 1 & 8  & 0.81 & 1 & 0 & 0 & 0 & 0 & 8  \\
9  & 1477.148868 & 1 & 0 & 1.2710D-01 & 1 & 0 & 0 & 0 & 4 & 0 & 0 & 1 & 1  & 0 & 0 & 1 & 9  & 0.84 & 0 & 0 & 0 & 0 & 0 & 16 \\
10 & 1679.479144 & 1 & 0 & 5.6793D-01 & 1 & 0 & 1 & 0 & 1 & 0 & 0 & 1 & 1  & 0 & 0 & 1 & 10 & 0.42 & 0 & 0 & 0 & 1 & 0 & 4  \\
11 & 1714.839900 & 1 & 0 & 1.5294D+01 & 1 & 0 & 0 & 2 & 0 & 0 & 0 & 1 & 1  & 0 & 0 & 1 & 11 & 0.80 & 2 & 0 & 0 & 0 & 0 & 0  \\
12 & 1862.945393 & 1 & 0 & 1.7244D-01 & 1 & 0 & 0 & 1 & 3 & 0 & 0 & 1 & 1  & 0 & 0 & 1 & 12 & 0.82 & 1 & 0 & 0 & 0 & 0 & 12 \\
13 & 1945.069104 & 1 & 0 & 1.1596D-01 & 1 & 0 & 0 & 0 & 5 & 0 & 0 & 1 & 1  & 0 & 0 & 1 & 13 & 0.74 & 0 & 0 & 0 & 0 & 0 & 20 \\
14 & 1967.624310 & 1 & 0 & 6.7322D-01 & 1 & 0 & 0 & 2 & 1 & 0 & 0 & 1 & 1  & 0 & 0 & 1 & 14 & 0.77 & 2 & 0 & 0 & 0 & 0 & 4  \\
15 & 1973.269531 & 1 & 0 & 2.8904D-01 & 1 & 0 & 1 & 0 & 2 & 0 & 0 & 1 & 1  & 0 & 0 & 1 & 15 & 0.39 & 0 & 0 & 0 & 1 & 0 & 8  \\
16 & 2240.375911 & 1 & 0 & 4.8422D+00 & 1 & 0 & 1 & 1 & 0 & 0 & 0 & 1 & 1  & 0 & 0 & 1 & 16 & 0.35 & 1 & 0 & 0 & 1 & 0 & 0  \\
17 & 2280.037817 & 1 & 0 & 2.8806D-01 & 1 & 0 & 0 & 2 & 2 & 0 & 0 & 1 & 1  & 0 & 0 & 1 & 17 & 0.72 & 2 & 0 & 0 & 0 & 0 & 8 \\
\hline
\end{tabular}
\end{center}
\noindent
\Dh{2}(M) represention: 1 is $A_g$, 2 is $A_u$, 3
is $B_{1g}$, 4 is $B_{1u}$, 5 is $B_{2g}$, 6 is $B_{2u}$, 7 is $B_{3g}$ and 8 is $B_{3u}$ \\
$I$:   State counting number; \\
$\tilde{E}$: State term energy in \cm; \\
$g$: State degeneracy; \\
$J$:   State rotational quantum number; \\
$L$:  Lifetime (s) \\
$\Gamma_{\rm tot}$: Total symmetry in \Dh{2}(M) ($A_g$, $A_u$, $B_{1g}$ and $B_{1u}$ only)  \\
$v_1$,$v_2$,$v_3$,$v_5$,$v_6$:  Normal mode vibrational quantum
numbers; \\
$n$:  torsional mode excitation; \\
$\tau$:  torsional symmetry; \\
$\Gamma_{\rm vib}$: Symmetry of vibrational contribution in \Dh{2}(M); \\
$K$:   State projection of the rotational quantum number; \\
$p$:   Rotational parity; \\
$\Gamma_{\rm rot}$: Symmetry of rotational contribution in \Dh{2}(M);  \\
$I_{J,\Gamma}$: State number in $J,\Gamma$ block; \\
$|C_i^{2}|$:  Largest coefficient used in the assignment;\\
$n_1 - n_6$:  TROVE vibrational quantum numbers.\\
\end{table}

\begin{table}
\caption{Extracts from the \hhoo\ transitions file.
The full table is available at
\protect\url{http://cdsarc.u-strasbg.fr/cgi-bin/VizieR?-source=J/MNRAS/}}
\label{tab:trans}
\begin{center}
\begin{tabular}{rrr}
\hline
       \multicolumn{1}{c}{$f$}  &  \multicolumn{1}{c}{$i$} &
\multicolumn{1}{c}{$A_{fi}$}\\
\hline
     4893324  &    4865894 & 2.5558e-33 \\
     6064947   &   6004042 & 5.0185e-24 \\
      978337   &    813317 & 2.3963e-29 \\
     5920140   &   5983274 & 9.1885e-25 \\
     4838267   &   4920666 & 1.5740e-25 \\
     2328674   &   2173645 & 7.2614e-33 \\
\hline
\end{tabular}

\noindent
 $f$: Upper  state counting number;\\
$i$:  Lower state counting number; \\
$A_{fi}$:  Einstein-$A$ coefficient in s$^{-1}$.

\end{center}
\end{table}

\begin{figure}
\centering
\epsfxsize=14.0cm \epsfbox{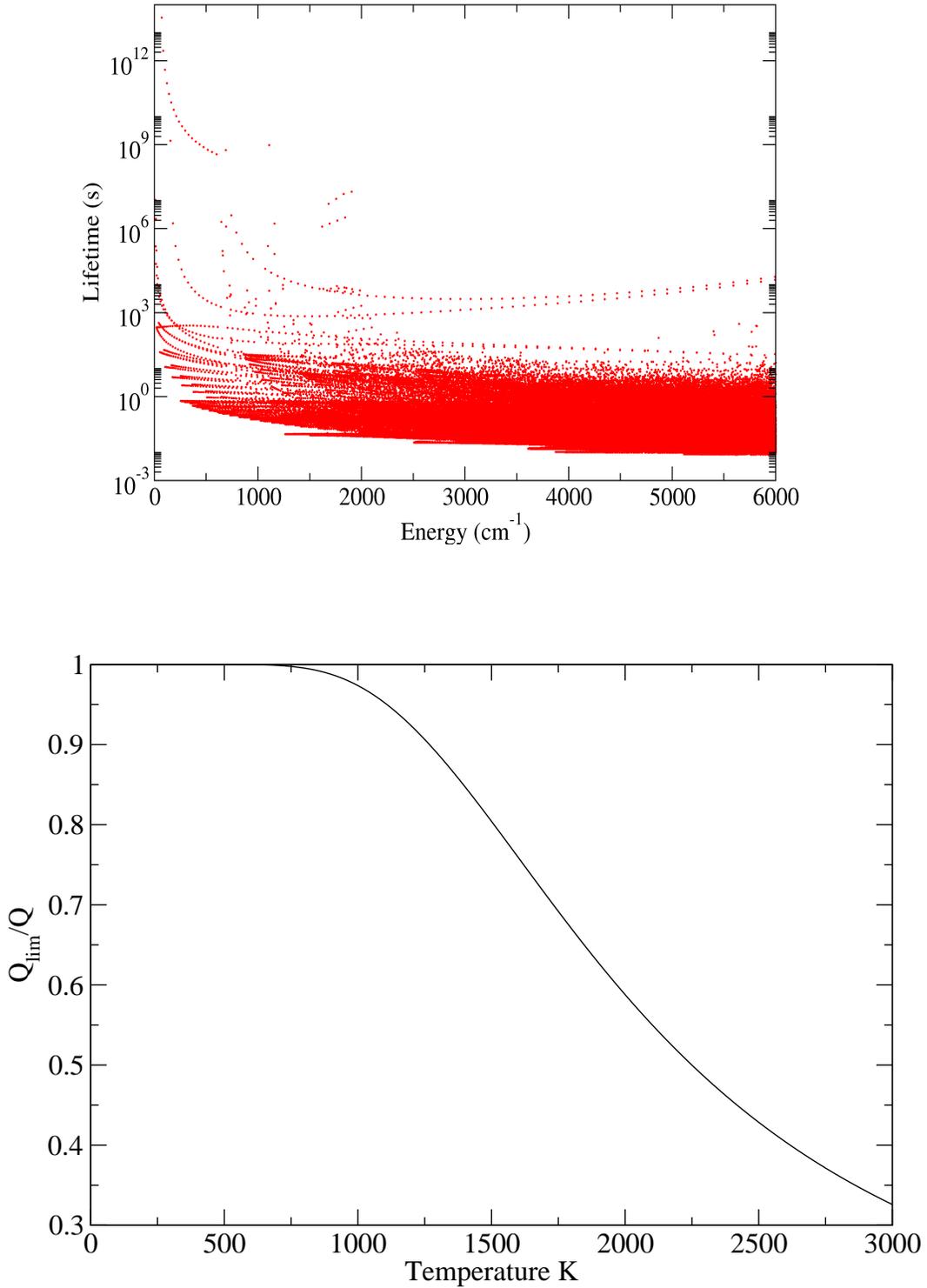}
\caption{\(Q_{\rm lim}\)/\(Q\) against temperature where \(Q_{\rm
limit}\) is the partition function
computed using only energy levels below our lower state threshold of 6000
cm\(^{-1}\) and $Q$ is our estimate of the full partition function.}
\label{fig:part_plot_limit}
\end{figure}

An estimate of the temperature applicability of the line list can be performed
by checking the partition function convergence of Eq.~(\ref{eq:part}), which is
computed via explicit summation \citep{jt263}. The convergence can be measured by computing $(Q_{J} - Q_{J-1})/{Q_J}$
where $Q_{J}$ is the partition function for all energy levels up to rotational excitation $J$.
For 296 K, the partition function converges to 0.001\% at $J=37$ which matches
the room temperature line list's $J$ limit of $J=40$.
At higher temperatures, it is well-converged up to at least 1500 K
where the estimated error is only 0.2~\% at $J=85$.
This can be attributed to the good coverage of $J$ states
computed that contribute to the overall population.
A second partition $Q_{lim}$ can be evaluated by only including states that
fall below the $hc$ 6\,000 \icm\ lower state energy limit of the line list and
compared against $Q$ by computing the ratio ${Q_{lim}}/{Q}$ to assess the
completeness of the full line list. Figure \ref{fig:part_plot_limit} shows that at
up to 800 K, the partition functions are essentially the same. At 1250~K
about 90\%\ of the population of states is represented by $Q_{lim}$ but this
falls to $\approx$80\% at 1500 K giving the upper temperature for which APTY is
reasonably complete as 1250 K. Usage of the line list at higher
temperatures runs the risk of losing opacity due to missing contributions. The
ratio ${Q_{lim}}/{Q}$ can be used to estimate this. However, the decomposition
of \hhoo\ means that it is unlikely to be an important species above 1000~K.
The partition function tabulated in steps of 1 K, alongside an associated
cooling function, is given as Supplementary Information to this article.

Table \ref{tab:part} compares our partition function against HITRAN;
we see that at temperatures less than 1000 K we agree better than 1$\%$. For
temperatures 1000 -- 1500~K, the APTY partition function is greater by 2--4\%\
suggesting that the explicit summation method gives higher, and probably better values,
than the more approximate method used by
HITRAN \citep{03FiGaGo.partfunc}. However, at 3000 K
APTY's $Q$ is 30\%\ lower than the HITRAN value which can be
attributed to the eigenvalue cutoff of
12\,000~\cm\ and $J=85$. Studies on ammonia and phosphine
have shown that considerably extended lists of energy levels
are required to get converged partition sums at these elevated temperatures \citep{jt571}.

\begin{table}
\caption{Comparisons of \hhoo\ partition functions as function of
temperature for this work those used in HITRAN \citep{03FiGaGo.partfunc}.}
\begin{center}
\begin{tabular}{rrr}
\hline\hline
$T$ / K    & APTY          & HITRAN         \\
\hline				            \\
75         & 895.506       & 894.866        \\
150        & 2~815.866     & 2~811.187      \\
255        & 7~360.598     & 7~336.856      \\
300        & 10~126.961    & 10~087.090     \\
500        & 31~246.17    & 30~990.11     \\
1000       & 232~439.8   & 226~152.5    \\
1500       & 1~031~673.6 & 993~983.8    \\
3000       & 21~847~680   & 15~151~254      \\
\hline
\end{tabular}
\end{center}
\label{tab:part}
\end{table}

\begin{figure}
    \centering
\epsfxsize=14.0cm \epsfbox{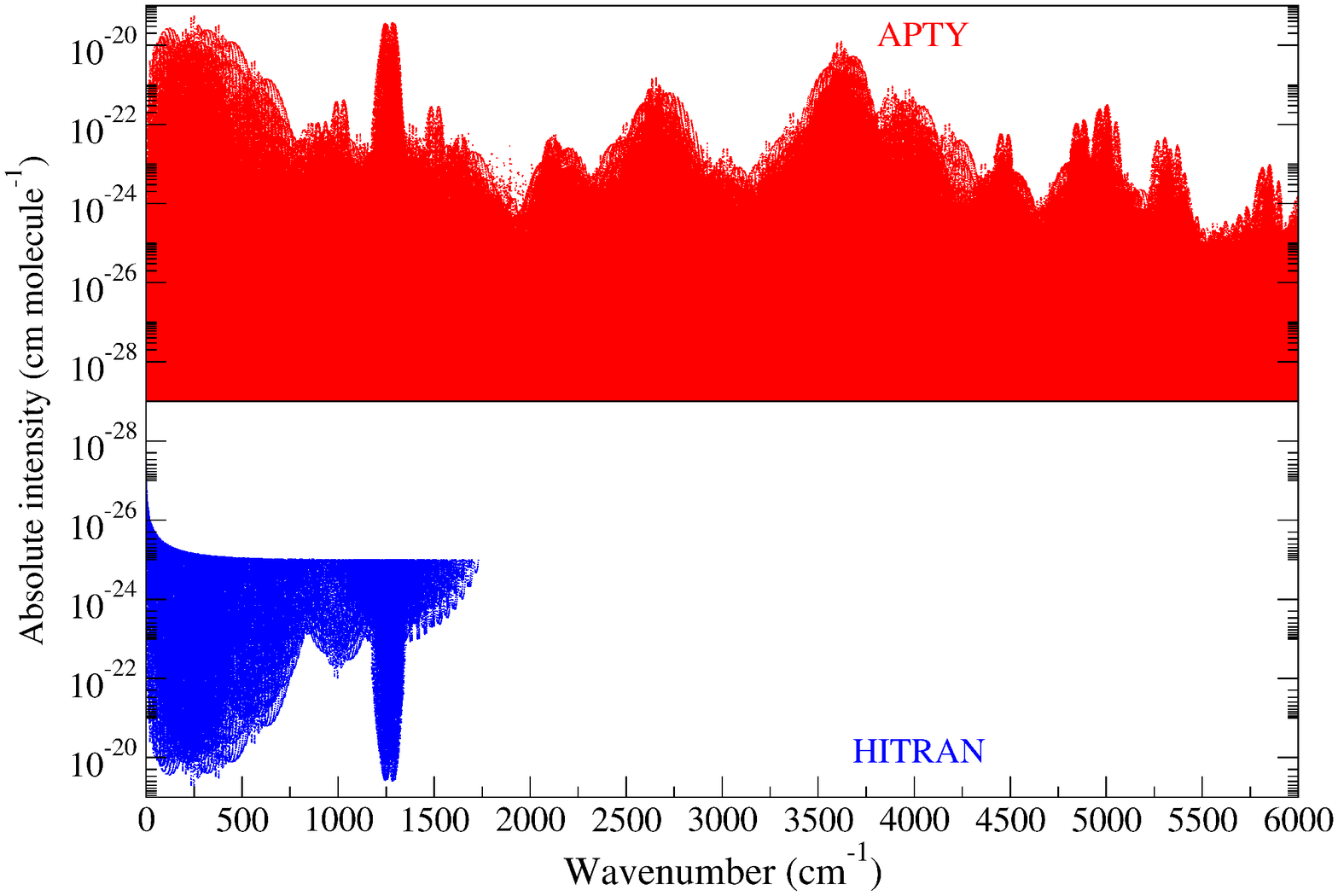}
\caption{Overview of our synthetic spectrum at $T=296$ K against HITRAN data  \citep{jt557}.}
    \label{fig:trove_hi_log}
\end{figure}

\begin{figure}
\centering
\epsfxsize=14.0cm \epsfbox{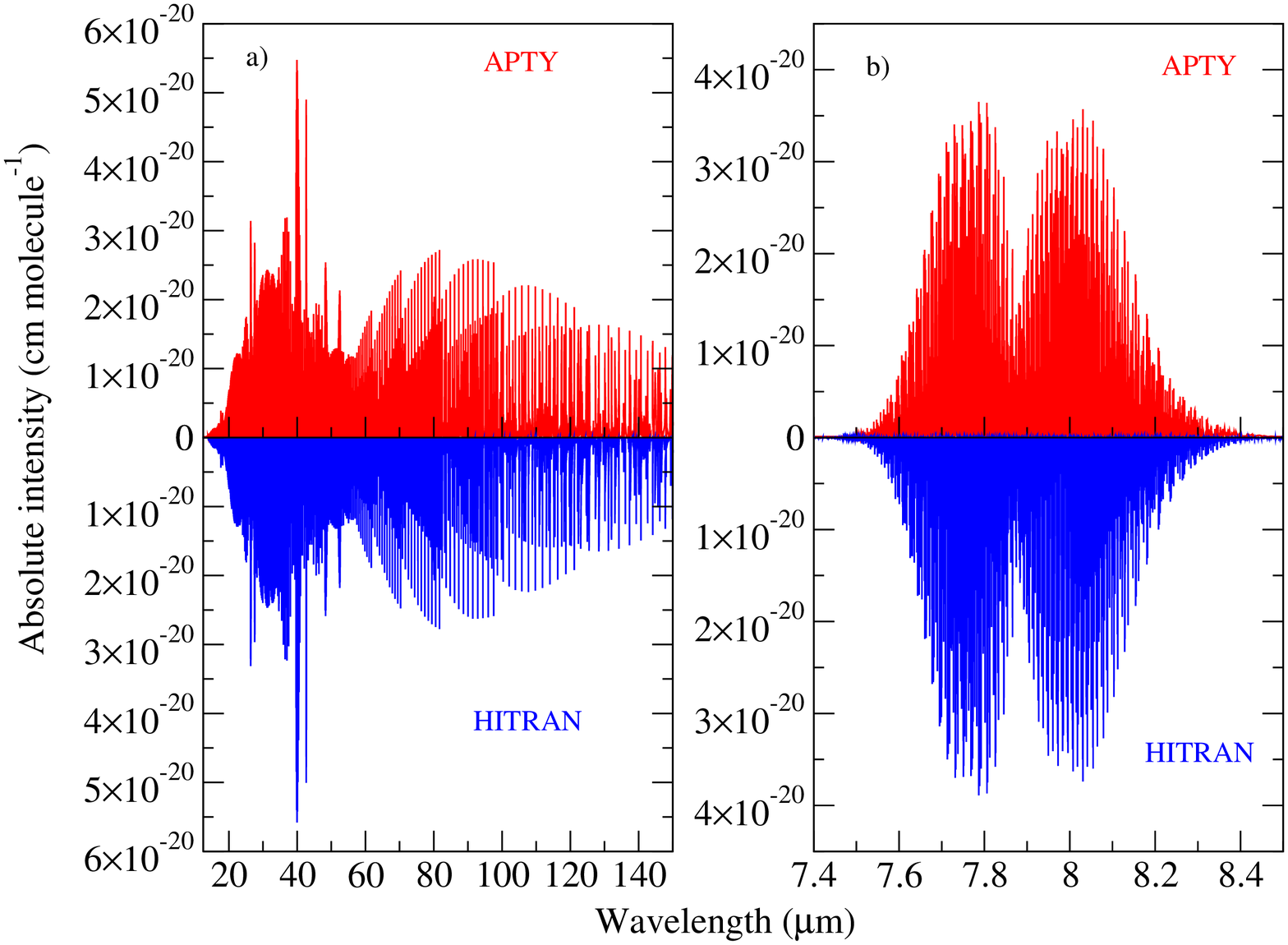}
\caption{The fundamental bands compared to the HITRAN database
\citep{jt557} at $T=296$ K: (a) Torsional and (b) $\nu_6$ bands.}
\label{fig:trove_2figs}
\end{figure}

Figure \ref{fig:trove_hi_log} is a simulated spectrum of the APTY line list
computed at $T=296$ K. This highlights the coverage and sheer number and density
of transitions available compared to the current edition of
the HITRAN database \citep{jt557}.  Figure \ref{fig:trove_2figs} compares our
results with specific
regions in the HITRAN database, the torsional and $\nu_{6}$ bands. Comparisons
of the two show excellent agreement in replicating both line position and
intensities.

\begin{figure}
\centering
\epsfxsize=14.0cm \epsfbox{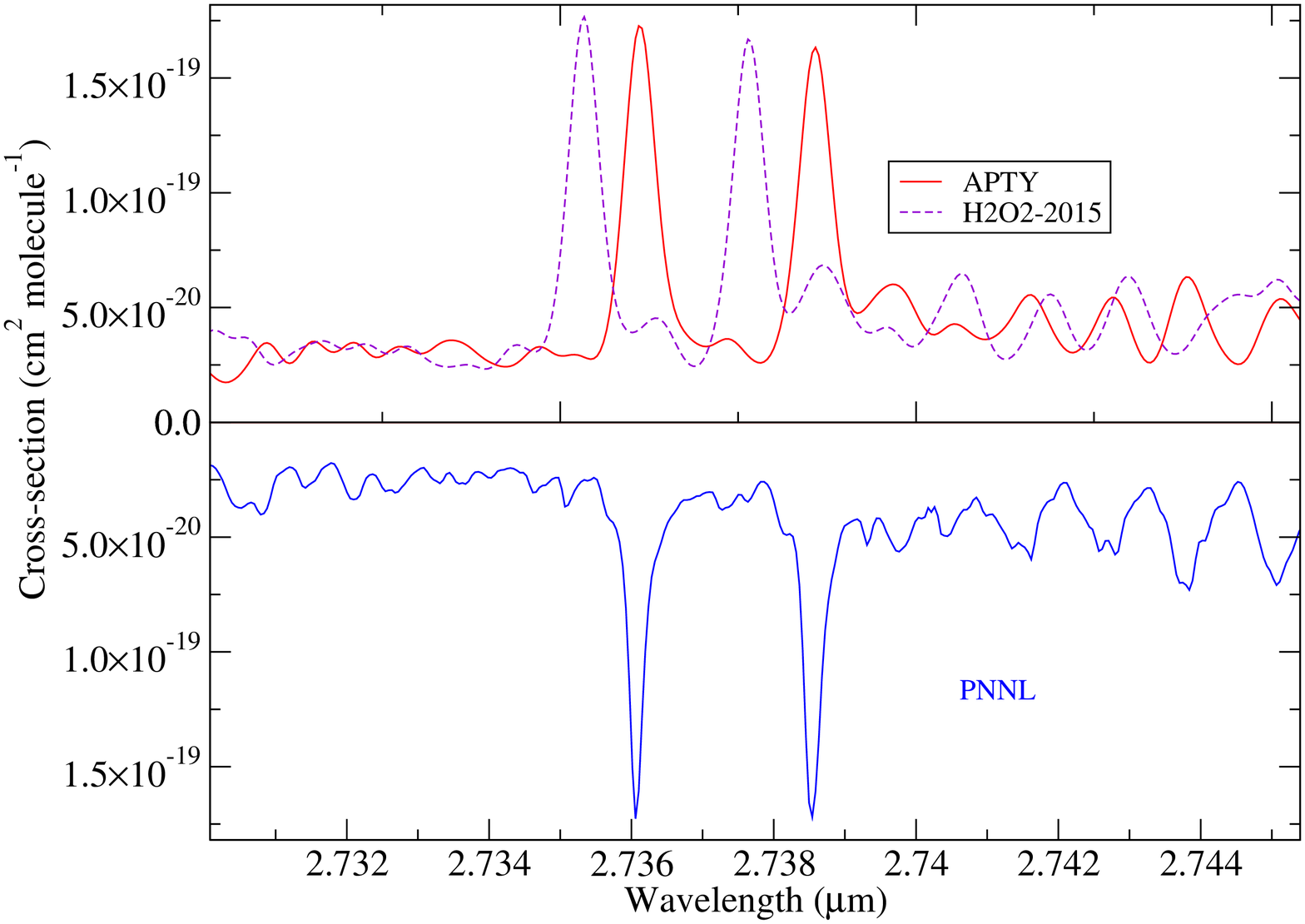}
\caption{Cross-section comparison of peaks in the $\nu_{1}$ and $\nu_{5}$ band
region with APTY (this work) and H2O2-2015 \citep{jt620} against PNNL-IR data at
323.15 K \citep{PNNL} with HWHM = 0.300 cm$^{-1}$ }
\label{fig:v1v5crossCompare}
\end{figure}

Our line list in the $\nu_{1}$ and $\nu_{5}$ band regions can be validated by simulating absorption cross
sections in the the 2.7~$\mu m$ region and comparing against PNNL-IR data \citep{PNNL}. The
structure and positions are in good agreement with the overall integrated
intensity for APTY in this region being 3\% stronger than PNNL. The improvement given by the
H2O2-2016 PES can be demonstrated by comparing with our previous room
temperature line list and with the PNNL-IR data; see Figure
\ref{fig:v1v5crossCompare}. The wavelengths of the largest two peaks in this band at 2.738~$\mu m$ and 2.736~$\mu m$
are correctly reproduced by APTY but are  shifted by about 0.001~$\mu m$ for
H2O2-2015, showing the improvement in this band due to use of the refined PES. Overall the
integrated cross-sections for this band differs only by 3\%\ from PNNL; indeed the entire
spectrum up to 6,000~\icm\ only differs by 3\%.

\begin{figure}
\centering
\epsfxsize=14.0cm \epsfbox{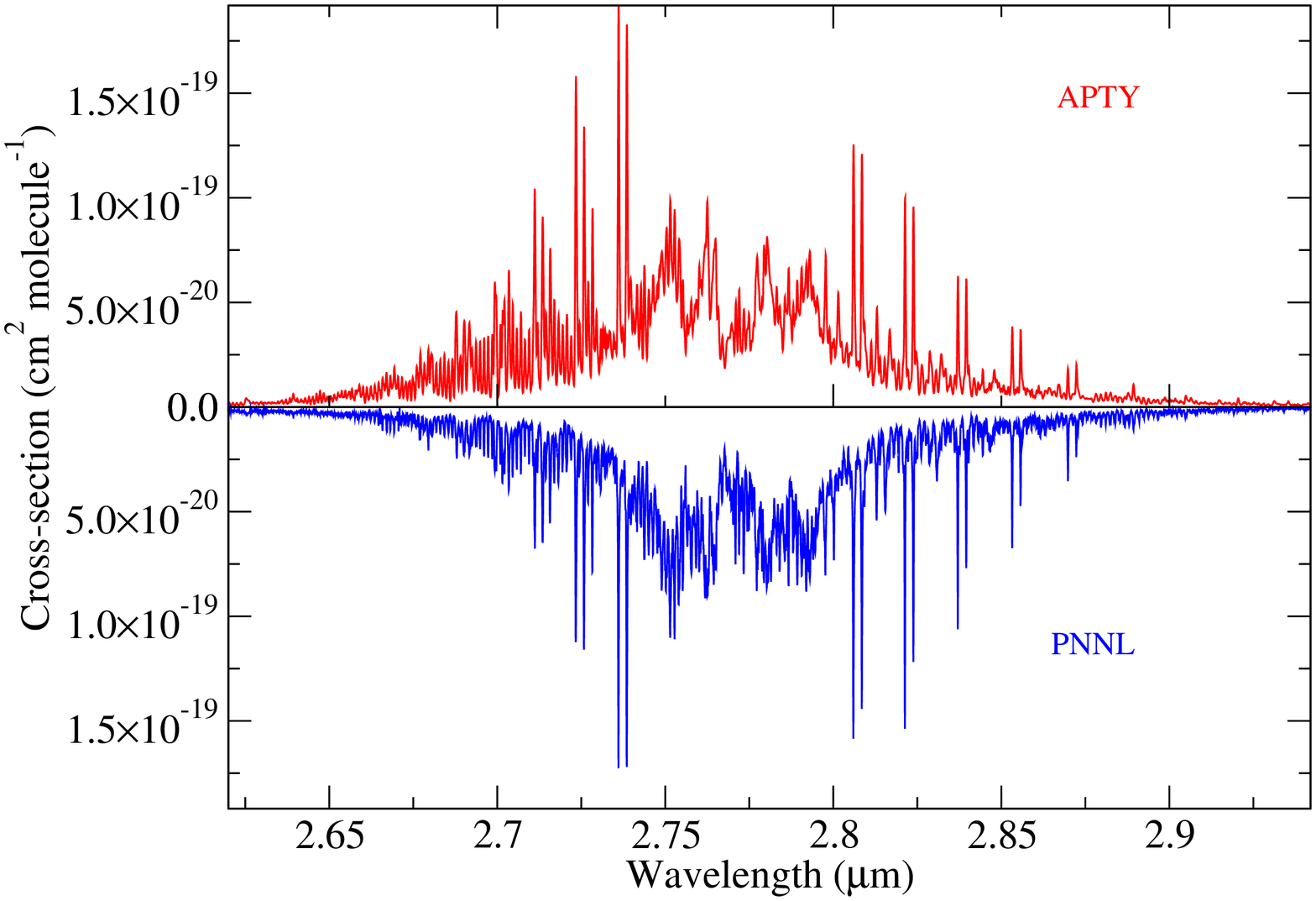}
\caption{The $\nu_{1}$ and $\nu_{5}$ band region with APTY against PNNL-IR data
at 323.15 K \citep{PNNL} with HWHM = 0.300 cm$^{-1}$ }
\label{fig:v1v5cross}
\end{figure}

Band intensities can be computed by explicit summation of all transitions
within a band and
compared against available data. Table \ref{tab:bandintens}
shows that for the limit available empirical band intensities we agree with most
regions to $\leq$7.3\% which is below the estimated experimental uncertainty of
$\approx\pm$10\%. Two discrepancies are with the $v_{6}$ band from
\citet{95PeVaFl.H2O2} and the $\nu_2+\nu_6$ band from \citet{09JoSaBu.H2O2}. The
former conflicts with other measurements due to \citet{09JoSaBu.H2O2} and
\citet{Klee1999154} where integrated absorption intensities were measured directly, whilst
the band intensities of \citet{95PeVaFl.H2O2} were obtained by summing a synthetic
spectrum of only 27~276 transitions. Our $\nu_2+\nu_6$
band intensity is 33.78\% weaker than the
experimentally derived value. \citet{09JoSaBu.H2O2} suggests that the assignment of this band is $\nu_2+\nu_6$ compared
to the $\nu_2+\nu_3+\nu_4$ assignment by \citet{50GiXXXX.H2O2}. This is based on a Q-branch peak observed at 2,658.62 \icm.
The assignments from APTY suggest that the peak observed is actually a convolution of Q-branches of the $(0,4) \rightarrow \nu_2+(0,4)$,
 $(0,4) \rightarrow \nu_2+(0,2)+ \nu_6 $, $(0,4) \rightarrow \nu_3+(4,4)$, $(0,4)\rightarrow\nu_3+(2,2)$, $(0,4)\rightarrow 2\nu_3+(2,3)$ and $(0,4)\rightarrow \nu_3+(2,2)+\nu_6$ transitions with an average separation between them at $\approx$0.02 ~\icm. Computing the band intensities of all of these bands in this region give an answer that agrees with value given by \citet{09JoSaBu.H2O2} to 3.97\%.

\begin{figure}
\centering
\epsfxsize=14.0cm \epsfbox{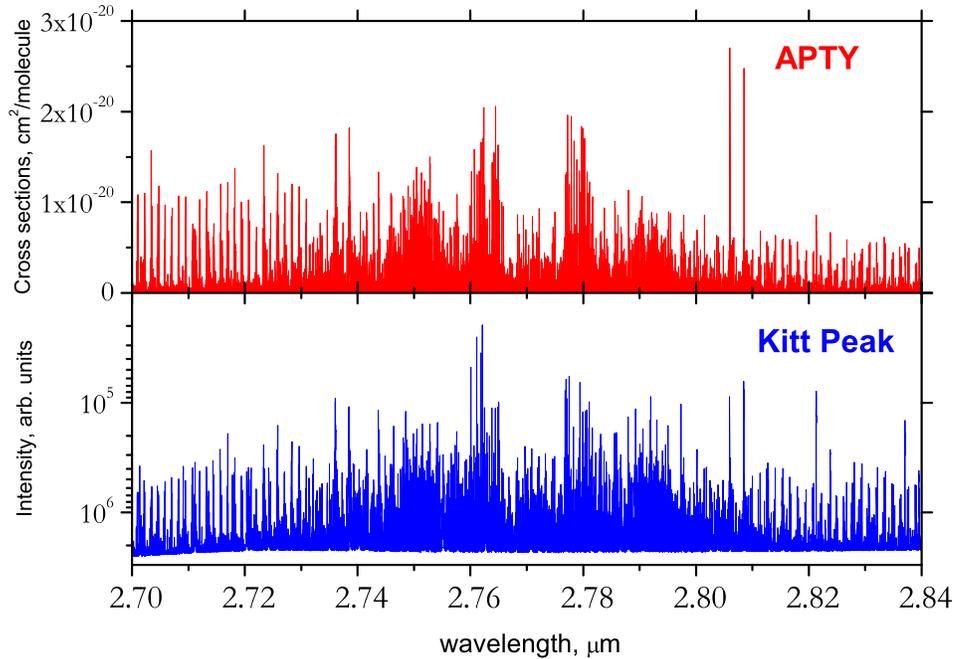}
\caption{ The 2.77 $\mu$m band of \hhoo\ at room temperature. Upper display: APTY cross-sections (296 K) generated using a Doppler profile; Lower display: an uncalibrated  Kitt Peak spectrum of \hhoo\ (Archive Name is 800628R0.002, Date is 28/06/1980;  Range is 1599.010271--6421.422198  \icm, Observer is Hunt;  1.77~m cell).
}
\label{fig:Kitt:Peak}
\end{figure}

The Kitt Peak Archive provides FTIR spectra of \hhoo\ covering the wavenumber region up to 6422~\icm, which is only partly assigned. Figure \ref{fig:Kitt:Peak} (lower display) shows an uncalibrated spectrum of \hhoo\ in the 1.78~$\mu$m region (800628R0.002) recorded by R. H. Hunt in 1980, which covers the $\nu_1$ and $\nu_5$ fundamental bands of the hydrogen peroxide. To the best our knowledge these two bands have not been spectroscopically analysed. The upper display of this figures presents our absorption spectrum at 296~K simulated using the Doppler line profile. Our synthetic spectrum resembles all the main features of the experimental data. We would like to encourage a spectroscopic analysis of the Kitt Peak \hhoo\ spectra in the IR and near-IR  regions currently not present in HITRAN. We believe that our theoretical line list with a capability of providing absolute intensities and quantum numbers can assist in the assignment of these spectra.


\begin{table}
 \caption{Band intensities, in 10\(^{-17}\) cm$^{-1}$/(molecule cm$^{-2}$).
}
\begin{center}
\begin{tabular}{llrllr}
\hline
Band & Frequency range (\icm) & Ref. & Obs & Calc & (O-C)/O (\%)      \\
\hline
Torsional                  & 0--1,427       &  \citet{96PeFlCA.H2O2}   &  4.0400  &  3.7450  &    7.3 \\
$\nu_3$                    & 750--1,100     & \citet{09JoSaBu.H2O2}    &  0.0157  &  0.0165  &  -5.43 \\
$\nu_6$                    & 1,135--1,393   & \citet{09JoSaBu.H2O2}    &  1.7458  &  1.7651  &  -1.10 \\
$\nu_6$     	           & 1,170--1,380   & \citet{Klee1999154}      &  1.8500  &  1.7633  &   4.68 \\
$\nu_6$                    & 1,170--1,380   & \citet{95PeVaFl.H2O2}    &  1.0030  &  1.7633  & -75.80 \\
$\nu_2+\nu_6$              & 2,300--2,900   & \citet{09JoSaBu.H2O2}    &  0.0830  &  0.055   &  33.78 \\
Multiple bands  	   & 2,300--2,900   & \citet{09JoSaBu.H2O2}    &  0.0830  &  0.0797  &   3.97 \\
$\nu_1$,$\nu_5$ region     & 3,300--3,800   & \citet{09JoSaBu.H2O2}    &  0.8356  &  0.8724  &  -4.40 \\
\hline
\end{tabular}
\end{center}
\label{tab:bandintens}
\end{table}

\begin{figure}
\centering
\epsfxsize=14.0cm \epsfbox{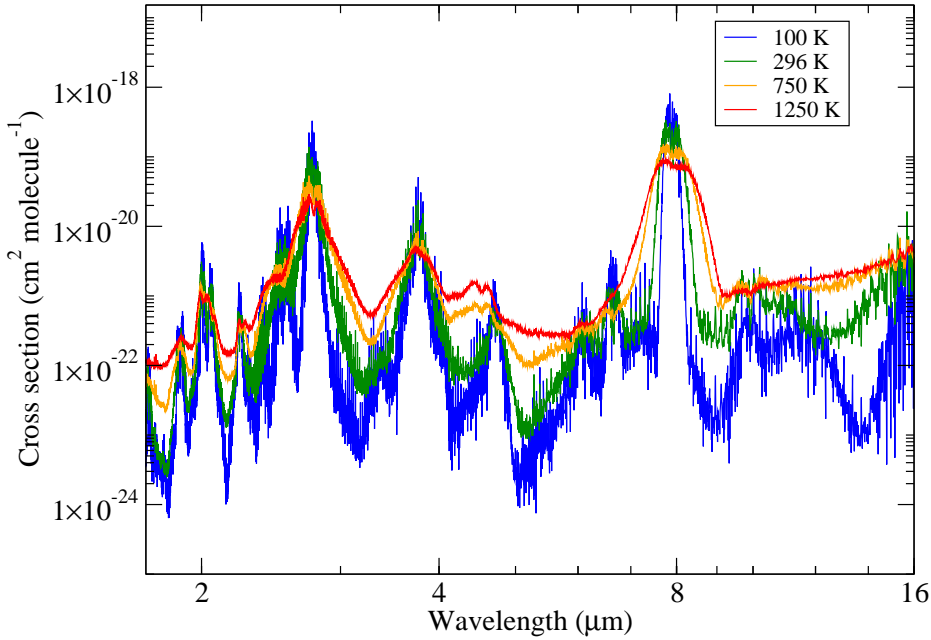}
\caption{ Cross-sections using a Doppler profile for the APTY line list as
a function of temperature.
The depth of the minima (windows)  decreases monotonically with temperature.
}
\label{fig:trove_dep}
\end{figure}

Figure
\ref{fig:trove_dep} presents
integrated absorption cross-sections computed  using a
Doppler profile \citep{jt542} for a range of temperatures.
The figure shows how the opacity changes with
increasing temperature.
We note the particularly dramatic effect raising the temperature has on
the absorption by \hhoo\
in the 13.7~$\mu$m region.
This smoothing in the overall spectra can only be
modelled if there is adequate coverage and population of rotationally and
vibrationally excited states.
We also note the strength of the OH stretch feature at about 2.75 $\mu$m;
these features are  absent from line databases such as HITRAN because of the
absence of assigned spectra in this region. Hopefully APTY can be used
to help analyse spectra in this region, as the BYTe NH$_3$ line list
is being used to analyse ammonia spectra \citep{jt616,jt633}.

\section{Conclusion}

The frequency and Einstein-$A$ coefficients of almost 20
billion transitions of hydrogen peroxide are computed. These
transitions cover wavelengths longer than 1.6
$\mu$m and include all rotational excitations up to \(J=85\), making
the line list applicable for temperatures up to 1250 K. The line list gives a
room-temperature spectrum in excellent agreement with available experimental
data and has good predictive ability for bands and line-positions not available
experimentally.
The new line list may be accessed via \url{www.exomol.com} or
\url{http://cdsarc.u-strasbg.fr/viz-bin/qcat?J/MNRAS/}. The cross-sections of
\hhoo\ can be also generated at \url{www.exomol.com} as described by
\citet{jt631}.

\section{Acknowledgements}
This work was supported by the ERC under the Advanced Investigator
Project 267219 and made use of the DiRAC@Darwin, DiRAC@COSMOS HPC cluster and
Emerald CfI cluster. DiRAC is the UK HPC facility for
particle physics, astrophysics and cosmology and is supported by STFC and BIS.
The authors would like to acknowledge the work presented here made use of the
EMERALD High Performance Computing facility provided via the Centre for
Innovation (CfI). The CfI is formed from the universities of Bristol, Oxford,
Southampton and UCL in partnership with STFC Rutherford Appleton Laboratory. We
would also like to thank nVidia for supplying our group with a Tesla K40 GPU for
usage in intensity and cross-section computation. AFA would
also like to thank Dr. Faris N. Al-Refaie, Lamya Ali, Sarfraz Ahmed Aziz, and
Rory and Annie Gleeson for their support.

\bibliographystyle{mn2e}

\label{lastpage}

\end{document}